\documentclass[review]{elsarticle}

\usepackage{lineno,hyperref}

\usepackage{color}
\usepackage{graphicx}
\usepackage{float}
\graphicspath{ {./images/} }

\begin{document}

\begin{frontmatter}

\title{Modeling of Lifeline Infrastructure Restoration Using Empirical Quantitative Data}



\author[mymainaddress]{Matthew Martell}
\ead{marte292@uw.edu}


\author[mysecondaryaddress]{Scott B. Miles}
\ead{milessb@uw.edu}

\author[mymainaddress]{Youngjun Choe\corref{mycorrespondingauthor}}
\cortext[mycorrespondingauthor]{Corresponding author}
\ead{ychoe@uw.edu}

\address[mymainaddress]{Department of Industrial \& Systems Engineering, University of Washington, Seattle}
\address[mysecondaryaddress]{Department of Human Centered Design \& Engineering, University of Washington, Seattle}

\begin{abstract}
Disaster recovery is widely regarded as the least understood phase of the disaster cycle. In particular, the literature around lifeline infrastructure restoration modeling frequently mentions the lack of empirical quantitative data available. Despite limitations, there is a growing body of research on modeling lifeline infrastructure restoration, often developed using empirical quantitative data. This study reviews this body of literature and identifies the data collection and usage patterns present across modeling approaches to inform future efforts using empirical quantitative data. We classify the modeling approaches into simulation, optimization, and statistical modeling. The number of publications in this domain has increased over time with the most rapid growth of statistical modeling. Electricity infrastructure restoration is most frequently modeled, followed by the restoration of multiple infrastructures, water infrastructure, and transportation infrastructure. Interdependency between multiple infrastructures is increasingly considered in recent literature. Researchers gather the data from a variety of sources, including collaborations with utility companies, national databases, and post-event damage and restoration reports. This study provides discussion and recommendations around data usage practices within the lifeline restoration modeling field. Following the recommendations would facilitate the development of a community of practice around restoration modeling and provide greater opportunities for future data sharing.
\end{abstract}


\end{frontmatter}

\section{Introduction}

Recovery from disasters is widely considered the least understood phase of the disaster cycle \cite{Miles2019,susdisrec}. Disaster recovery is a broad term that has many facets including social, economic, built and natural environments. It is largely accepted to imply bringing each of these facets back to or better than pre-disaster levels \cite{ChangStephanieE.2010Udra,Kates1977,Lindell2013}. A subsection of disaster recovery research is lifeline restoration modeling. Restoration refers to the patching up of essential services to help facilitate longer-term recovery \cite{Kates1977,Loggins2019CMtR}. Lifelines are a subset of critical infrastructures vital for communities to operate \cite{ppd21}, namely electricity, natural gas, telecommunication, transportation, water, wastewater and liquid fuel \cite{NISTgcr}. Understanding how these systems are restored allows for more informed community resilience planning efforts \cite{NISTcommresil,TDoRourke}. We can better understand lifeline restoration processes through modeling.  

The lack of, or perceived lack of, empirical data is one of the primary challenges for the growth of the lifeline restoration modeling field \cite{ChangStephanieE.2010Udra,MilesScottB.2006MCRf}. Ouyang \cite{OuyangMin2014Review} identifies difficult to access data and lack of precise data as key problems for modeling lifeline systems. Lifeline modeling requires a lot of data, frequently including system topologies, component geographical locations, and emergency procedures used by the lifeline system's owners. Data access is difficult for reasons such as antitrust laws, confidentiality, and privacy. Rinaldi et al. \cite{RinaldiS.M2001Iuaa} also identifies the volume of data required to model lifeline systems as a major challenge in the field. Ouyang \cite{OuyangMin2014Review} calls for a uniform data collection method to remedy data issues, while Miles et al. \cite{Miles2019} calls for a community of practice to develop around the broader field of disaster recovery modeling, including development of shared data sets. 

The need for a consistent approach to handling data in lifeline restoration modeling is apparent. To propose such an approach, it is necessary to understand the history of data usage in the field. This paper reviews the usage of empirical quantitative data to model lifeline infrastructure restoration. Section~\ref{sec:high-level-trend} discusses high-level trends seen in the literature and the literature search methods used. Section~\ref{sec:model} breaks down the literature by modeling approach for an in-depth look at how various approaches utilize empirical quantitative data. Section~\ref{sec:discussion} discusses topics related to lifeline restoration modeling such as model validation and testing methods, modeling interdependent systems, benchmarking testbeds and  unique data sources. Section~\ref{sec:reproducibility} introduces a consistent methodology for handling data in the lifeline restoration modeling domain to inform a standard for reproducible research and shared data sources.   

\section{High-Level Trends}\label{sec:high-level-trend}

We identified initial papers to include in the review by searching Web of Science for recent publications using the keywords lifeline, infrastructure, restoration, disaster, recovery, and data. Using the initial papers, we found older publications using backwards snowballing. Backwards snowballing is a technique for searching the literature by proceeding backwards in time through references of known papers to find older sources on a topic \cite{reviewpaperexample}. In total, we identified 54 papers for this study. As there are inconsistencies in the literature regarding the usage of key terminologies such as restoration, recovery, and response \cite{Miles2019}, we do not claim that this list is exhaustive. However, we believe it to be representative of the literature in the field. 

The literature analyzed for this paper is a subset of disaster recovery and modeling literature. It is useful to identify some excluded papers to illustrate the boundary of the reviewed literature. Nejat and Ghosh \cite{Nejat2016} use empirical data to model housing recovery, but their work is excluded from this review since housing is not considered a lifeline. Similarly, papers that model greater community recovery, or other non-lifeline sectors, are not included in this study \cite{MilesScottB2011RACB,BarkerKash2009Auie}. Works that collect restoration data without building a restoration model such as Nojima and Maruyama \cite{nojima2016comparison} are also not included. Additionally, papers that work with more qualitative data, such as expert judgements \cite{ChangStephanieE.2014TDCC}, are not included. A large body of literature omitted from this study concerns the power service restoration problem defined by the Institute of Electrical and Electronics Engineers (IEEE), as they use a specific technical definition. The problem is also known as the Fault Isolation and Service Restoration problem. Solutions to this problem try to find the fastest way to isolate a fault in the power distribution network, while minimizing the number of healthy out-of-service areas \cite{MarquesLeandroTolomeu2018SRWP}. There are reviews of the literature in this area including \'{C}ur\v{c}i\'{c} et al.\cite{ĆurčićS1995Epdn} and Liu et al. \cite{LIUYutian2016Psra}, so we refer readers to these papers for more information on this problem. Many papers in this domain use electricity infrastructure data, so they are a potentially valuable data source. Making exclusions of the above types allows us to keep our scope narrow while still having a significant body of research to review. 

\begin{figure}[!h]
\caption{Number of publications over time on modeling lifeline restoration using empirical quantitative data.}
\centering
\includegraphics[width=1\textwidth]{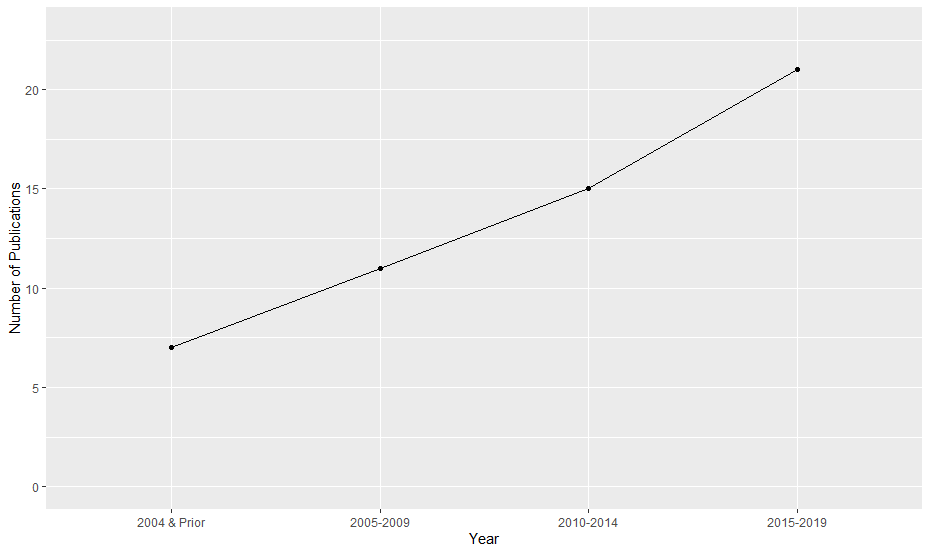}
\label{time}
\end{figure}

An initial finding from this review is that lifeline restoration modeling is a growing field. Figure~\ref{time} shows the marked increase in publications over time. 

\begin{figure}[!h]
\caption{Number of publications over time by modeling approach.}
\centering
\includegraphics[width=1\textwidth]{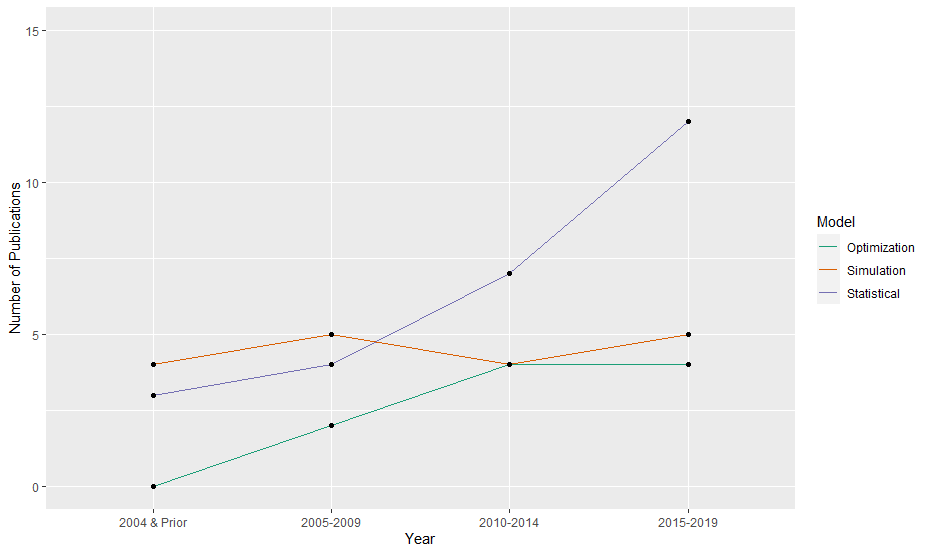}
\label{modtime}
\end{figure}

The sharp increase in publications over the last ten years (2010-2019) coincides with the proliferation of statistical models of lifeline restoration. Figure~\ref{modtime} shows the change in modeling approaches over time. Statistical modeling has grown markedly in the last ten years compared to other modeling approaches. This trend may be related to changes in the amount of available data and in what data is being used. The availability of outage/restoration data has likely increased with the increasing number of weather-induced disasters \cite{climatecentral}. This increase is in contrast with the availability of lifeline-specific data (e.g., topology of a networked system) typically used by simulation and optimization approaches. This type of data has not experienced the same trend in accessibility as outage/restoration data since it requires collaboration with utility companies. While statistical models can use publicly available community attributes, such as demographic information or economic data as predictors for outage duration, optimization or simulation approaches require some amount of lifeline-specific data to model the restoration process. Thus, the growth in statistical models is only natural. The data usage patterns of each modeling approach are discussed in more detail in Section \ref{sec:model}. 

\begin{figure}[!h]
\caption{Breakdown of the reviewed literature by hazard type. `Other Wind' includes ice storms and tornadoes.}
\centering
\includegraphics[width=1\textwidth]{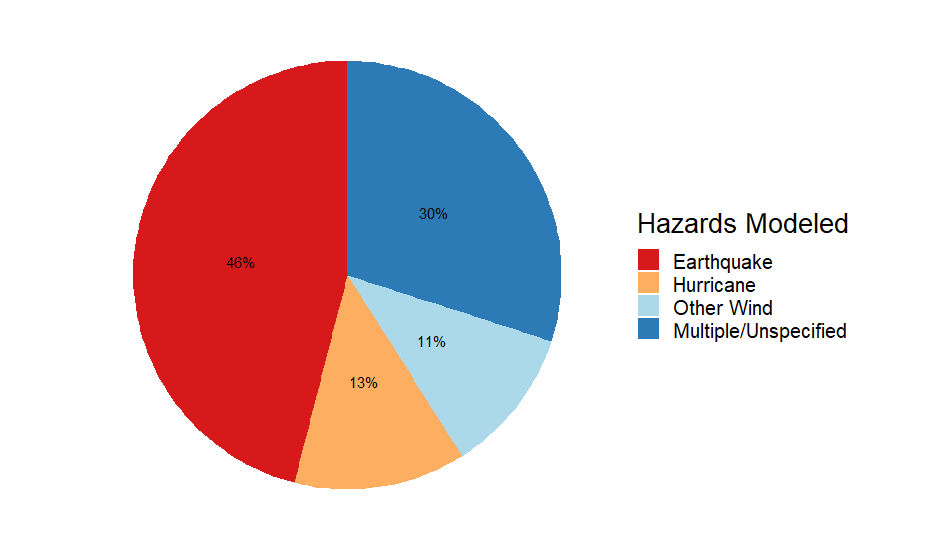}
\label{haz}
\end{figure}

Data availability is not the only factor that affects modeling decisions. Earthquake hazard research is a historically more organized and well-funded research domain than other hazards research. This is exemplified by major earthquake engineering research centers such as the Mid-America Earthquake Center, Pacific Earthquake Engineering Research Center, and the Multidisciplinary Center for Earthquake Engineering Research (MCEER). MCEER, formerly known as the National Center for Earthquake Engineering Research, alone produced hundreds of publications, some of which involve lifeline restoration modeling \cite{MCEER}. Two of the most extensive past restoration modeling and data collection efforts are MCEER projects that involved collaborations with the Los Angeles Department of Water and Power (LADWP) and Memphis Light, Gas and Water Division (MLGW). Both partnerships resulted in multiple publications, so earthquake-related models are heavily represented in the literature as seen in Figure~\ref{haz}. Another insight from Figure~\ref{haz} is that there is a large body of literature that assumes an initial damaged state without specifying a hazard type, or considers multiple hazards, to make those models more generalizable. 

\section{Modeling Approaches}\label{sec:model}

\begin{figure}[!h]
\caption{Breakdown of the reviewed literature by modeling approach.}
\centering
\includegraphics[width=1\textwidth]{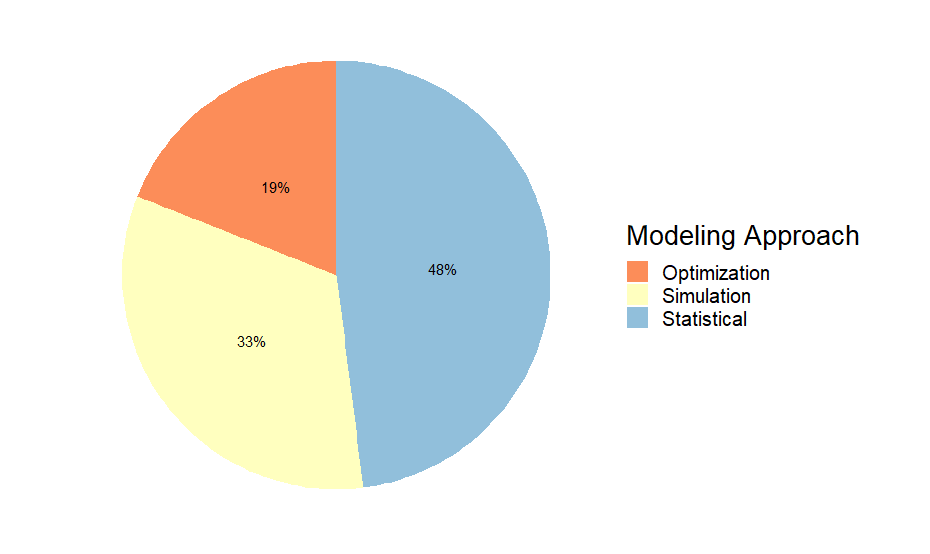}
\label{mod}
\end{figure}

We separate lifeline restoration modeling into three categories for our analysis: optimization, simulation and statistical modeling. While these categories are broad, there are still clear differences in data usage between them. These differences are enough to facilitate our discussion of data management practices, so further model categorization is unnecessary. This section discusses each modeling approach and the common data-usage practices within them. Statistical modeling approaches are the most common, followed by simulation, and then optimization 
(see Figure~\ref{mod}).

\begin{figure}[!h]
\caption{Breakdown of the reviewed literature by modeling approach and lifeline modeled.}
\centering
\includegraphics[width=1\textwidth]{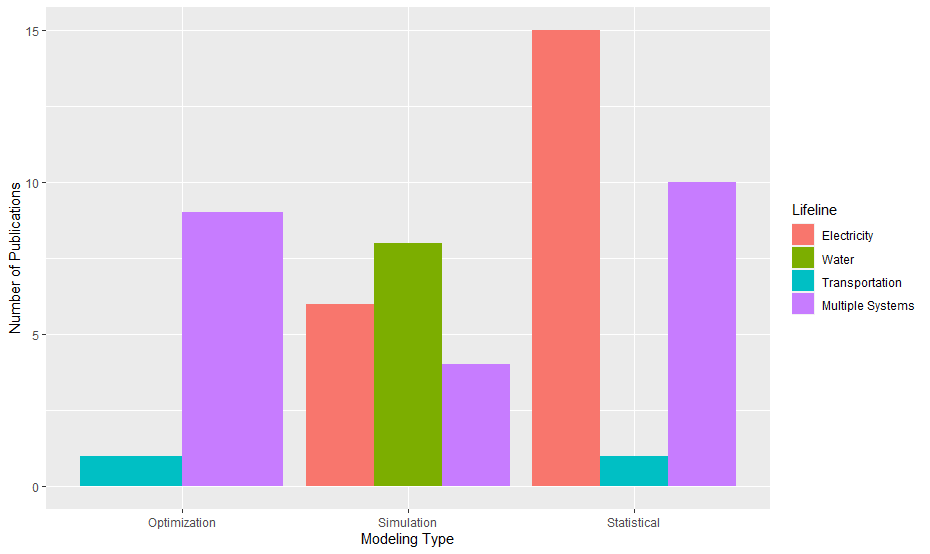}
\label{modlife}
\end{figure}

\begin{figure}[!h]
\caption{Breakdown of the reviewed literature by modeling approach and hazard modeled.}
\centering
\includegraphics[width=1\textwidth]{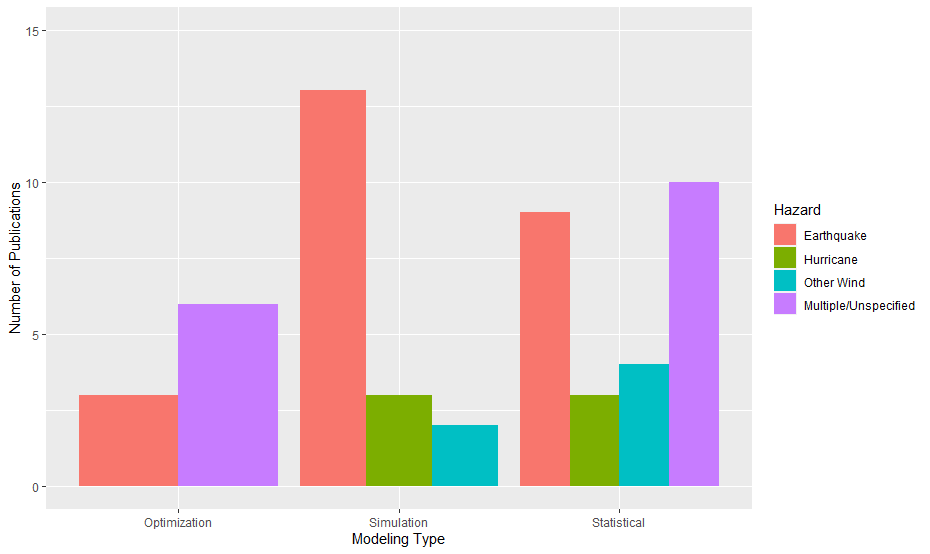}
\label{modhaz}
\end{figure}

There are clear connections between the modeling approaches and the types of data used. Optimization models most often consider multiple lifeline systems and hazard types, while statistical models are typically linked to electricity restoration and simulation models to earthquakes (see Figures~\ref{modlife} and \ref{modhaz}). Optimization models are procedural and emphasize generalizability, so they use data sets that represent multiple systems and hazards. Statistical approaches to modeling electricity restoration are common because power outage data are more common than outage data for other lifelines. Electricity restoration models are often constructed using outage data and any data that can be used as a predictor (e.g. electricity system features, hazard strength or socioeconomic data about the surrounding community). Simulation modeling of post-earthquake restoration is common because of the MCEER research program. The long-term MCEER partnership with the Los Angeles Department of Water and Power (LADWP) yielded high-resolution simulation models of post-earthquake restoration which in turn resulted in multiple publications. 

\subsection{Simulation}

\subsubsection{Overview}
Simulation models have the longest history of any method in the lifeline restoration modeling domain, dating back to the 1980s \cite{Isumi1985, IsoyamaRyoji1981PPEO}.
Simulation modeling is the second most common modeling approach in the literature reviewed, with statistical modeling taking the top spot in the last 10 years. In terms of data usage, simulation models are typically based on lifeline-specific data such as a connected graph representation of the system, individual component repair times, and available repair resources (e.g., maintenance crews). Two of the largest simulation data sets stems from the work modeling the restoration of water and power systems in Los Angeles, CA \cite{Cagnan2007,Cagnan2006,Xu2007,Tabucchi2010,Brink2012,Brink2009, mceer080008}. 

\subsubsection{Data Set Features}
Data for simulation models come from many different sources. In spite of this, there is a high level of overlap in the features of the data sets. Every simulation-based paper reviewed used lifeline infrastructure data in some capacity. Lifeline systems are commonly represented as connected graphs \cite{  IsoyamaRyoji1981PPEO,Cagnan2006, Ramachandran2015, Choi2018}. Component failure rates are frequently obtained from other works \cite{IsoyamaRyoji1981PPEO, Brink2009,Sun2015,  BrownR.E1997Dsra}. Another common data set feature is repair crew information such as repair rate/efficiency and number of crews \cite{Cagnan2007,Cagnan2006,Xu2007,Tabucchi2010,Brink2012,Brink2009, Luna2011}. Lastly, information about restoration from an outage event is primarily used for model validation and testing. Validation and testing methods are discussed more in Section \ref{sec:discussion}.

\subsubsection{Notable Data Sets}

This subsection, and the corresponding subsection for the other modeling approaches, focuses on highlighting some data sets from the lifeline restoration modeling literature. These data sets are highlighted because of their size or unique features.. 

Two of the largest data sets used for simulation modeling in the reviewed literature are those used to model the restoration of the LADWP systems \cite{  Cagnan2007, Cagnan2006, Xu2007,Tabucchi2010,Brink2012, Brink2009, mceer080008}. The data sets for these papers are the result of extensive collaboration with LADWP. The papers are from two separate projects, one for water restoration \cite{Tabucchi2010,Brink2012, Brink2009,   mceer080008} and one for power restoration \cite{Cagnan2007,Cagnan2006,  Xu2007}. The data sets include detailed network representations of the respective lifelines, locations of the various resources necessary for repair work, expected behavior of repair crews and the availability of each repair resource. Additionally, restoration and initial damage data from the 1994 Northridge earthquake serve as the basis for model validation.

Another notable data set is that used in one of the first papers modeling lifeline restoration \cite{Isumi1985}. The authors use a data set from Sendai city in Japan after the 1978 Off-Miyagi Earthquake. The data set includes census information on population and infrastructure information on gas, water and electricity systems, such as their respective layouts, characteristics and repair strategies, the number of available workers for each system, damage data for each system and time series outage data for each system. The simulation uses differential equations to describe the repair of damage over time according to a repair rate, where the repair rate was based on worker availability and productivity data. Time series restoration data is used to validate the model.

Sun et al. \cite{Sun2015} uses a similar modeling approach to Isumi et al. \cite{Isumi1985}. The model uses a simplified version of the IEEE 118 Bus Test Case, representing a midwestern U.S. electrical system from 1962. The authors produce a simplified model of the community supplied by the system by classifying buildings as residential, industrial or critical facilities. They use fragility functions found in previous works and HAZUS to model damage to the electricity system. 

He and Cha \cite{Xian2018} use a hypothetical study region to demonstrate their model before applying the model to a data set from Galveston, TX after Hurricane Ike. The data set describing the Galveston area includes locations of facilities for power, water and telecommunications lifelines, city zoning data and maximum wind speeds for different lifeline facility locations. The authors use accident report data from the National Transportation Safety Board and news media to estimate infrastructure dependencies. The restoration time of the electricity system is also used for validation.

Luna et al. \cite{Luna2011} researches water supply system restoration from earthquakes using discrete event simulation and a colored Petri nets approach. They use the data set of Isoyama and Katayama \cite{IsoyamaRyoji1981PPEO}. The data set includes the network representation of the trunk water supply system for Tokyo, damage probabilities for system components, repair crews, trucks, replacement pipes and excavators. The authors compare their model against \cite{IsoyamaRyoji1981PPEO}; however, they do not use baseline restoration data to test the model. 

\subsubsection{Data Sources}
A wide variety of data sources are used in simulation modeling studies, although some papers do not identify an original source for their data sets. Luna et al. \cite{Luna2011} and Brown et al. \cite{BrownR.E1997Dsra} use data sets from previous works for their models. Sun et al. \cite{Sun2015} uses an IEEE Bus Test Case for their network data as well as data from HAZUS and previous works for component fragility functions. Several papers in this area \cite{Cagnan2007,Cagnan2006,Xu2007, Tabucchi2010,Brink2012,    Brink2009, mceer080008} collaborate directly with LADWP and collect extensive data sets through interviews and reviewing emergency response plans. Other data sources include HAZUS, S\&P Global Platts (a provider of information for commodities markets), public utility data, government disaster reports and previous publications \cite{Ouyang2012A,OuyangMin2014Mhra, OuyangMin2015Raoi}. Isumi et al. \cite{Isumi1985} uses damage and restoration reports from local government and utility companies. Chang et al. \cite{Chang2002} makes use of a large data set collected in a previous effort \cite{Chang1996}. Lastly, Google Earth is an infrequent but inventive data source for identifying lifeline facility network structure \cite{Ramachandran2015,Xian2018}. 

\subsection{Optimization}
\subsubsection{Overview}

The purpose and therefore data usage of optimization modeling studies differ greatly from the other two modeling approaches. The purpose of an optimization model is typically to identify an efficient restoration sequence. In contrast, the purpose of simulation modeling is often to understand a restoration process in greater detail, while the purpose of statistical modeling is often to predict outage duration. Optimization models also distinguish themselves from other approaches by more frequently modeling interdependencies between lifelines through model constraints. 

From a data usage perspective, optimization models do not put as strong of an emphasis on using empirical data. This is in line with the typical purpose for optimization models compared to other approaches. Real-world data is not strictly necessary to prove a theoretical result such as optimality or show computation times. This is how data sets such as the one used by Lee et al. \cite{Lee2007} arise, where a realistic representation of several lifelines is generated using empirical quantitative and qualitative data together. 

\subsubsection{Data Set Features}
Optimization models are similar to simulation models in that they focus their modeling efforts on the lifeline systems and restoration processes themselves. This leads to data sets that take the form of connected graph representations of lifelines. These representations include location and capacity of supply nodes, node-arc lifeline interdependencies, flow capacities, flow costs, and repair costs.  

\subsubsection{Notable Data Sets}

Lee et al. \cite{Lee2007} is one of the more frequently cited optimization restoration modeling papers, and the data set they created is reused in multiple other papers \cite{Nurre2012, NurreSarahG.2014Inda, Cavdaroglu2013}. The authors use census data, data from the NYC Metropolitan Transit Authority, data from a local electric company and from Verizon to build a realistic representation of the lifelines in lower Manhattan. This representation includes physical layout, supplies, demands, capacities, interdependencies, as well as origin-destination information for the transportation and telecommunications networks. 

Nurre et al. \cite{Nurre2012} use the same data set for lower Manhattan as Lee et al. \cite{Lee2007}, in addition to collecting data about New Hanover County, NC. The New Hanover County data set includes representations of electricity systems, wastewater systems, and emergency supply chain infrastructures. This data set exists due to collaborations with the managers of the infrastructure systems, as well as the emergency manager for the county. All systems are represented as connected graphs; restoration strategies are implemented using the input of emergency and utility managers. Sharkey et al. \cite{SharkeyThomasC2015InrO} also uses this New Hanover County data set. Iloglu and Albert \cite{Iloglu2018} use a different data set from New Hanover County, representing the road network, locations of fire and rescue stations, and locations of demand for emergency services. 

In their papers, Gonz\'{a}lez et al. test their models on a data set representing Shelby County, Tennessee \cite{González2017,González2016}. It contains network representations of the power, water and gas systems of the county. This data set stems from an extensive partnership with a utility company, in this case MLGW. This partnership, like that with the LADWP, yielded one of the largest data sets on lifelines that has been used in many subsequent studies. It dates back to an MCEER project with many contributors, such as S. Chang and M. Shinozuka. This data set is discussed in more detail in Section~\ref{subsec:stat}.

Yan and Shih \cite{Yan2009} examine road network repair and disaster relief distribution. Their data set is from Taiwan after the 1999 Chi-Chi earthquake. It contains roadway-network information, emergency repair resources and commodity supplies and demands. Tuzun Aksu and Ozdamar \cite{TuzunAksu2014} likewise examine road network restoration of road networks of two districts in Istanbul, Turkey.

\subsubsection{Data Sources}

The data sources for optimization models are similar  to those of simulation models, but less varied. Collaboration with lifeline management organizations to get data is a common method \cite{Lee2007,Nurre2012,Yan2009}. Authors also consistently make use of data sets collected from prior studies \cite{Nurre2012,NurreSarahG.2014Inda,Cavdaroglu2013,SharkeyThomasC2015InrO,González2017,González2016}, frequently other lifeline restoration modeling efforts. There is less emphasis on data collection and usage than for other modeling approaches as the purpose of optimization models are frequently theoretical. Overall, optimization approaches use a similar, yet smaller set of data sources than simulation approaches.

\subsection{Statistical Models}\label{subsec:stat}

Statistical models are the most frequently used and most varied of the three modeling approaches. The goal of such a model is usually to generate a restoration time estimate (e.g. it will take 4 days for the lifeline to be 90\% functional), or a restoration probability (e.g. there is an 80\% probability the lifeline has 90\% functionality in 3 days). The statistical modeling approaches include curve fitting \cite{Park2006}, survival analysis \cite{Bessani2016,Davidson2017,Mojtahedi2017}, various machine learning techniques \cite{Nateghi2011,Mukherjee2018} and economic models \cite{Mackenzie2013}, among others.

With the widest variety of approaches, statistical models also encompass the widest variety of data set features and sources. A commonality amongst the statistical models is the use of lifeline restoration data used for model fitting and/or model validation and testing. Some larger data sets include power restoration after several hurricanes in the U.S. Gulf Coast region \cite{Nateghi2011,Nateghi2014} and a data set for power restoration after hurricanes and ice storms for three power companies covering North Carolina, South Carolina and Virginia in the U.S. \cite{Davidson2017,Liu2007,ReedDorothyA2008}. 

\subsubsection{Data Set Features}

A common feature of statistical modeling data sets is the use of restoration data from historical disaster events. Sometimes this takes the form of time series restoration data and other times a single data point representing X\% restoration for a particular geographic area. Lifeline data are also used in many studies.Common features for power system data sets include the number of poles, transformers, switches and lines in each grid cell of a spatial data layer \cite{Nateghi2011,Nateghi2014,Liu2007}. Other common data set features include hazard data such as wind speed, rainfall and ice accretion and geographic data such as land cover and soil depth \cite{Guikema2010, Davidson2017}. Several studies use socioeconomic data  \cite{Liu2007, Mitsova2018}, including demographics, population density and poverty rates. Other data set features include commodity trade data and climate data, such as mean annual precipitation \cite{ Mukherjee2018, Mackenzie2013,Guikema2010}. 

\subsubsection{Notable Data Sets}

In two papers, Nateghi et al. \cite{Nateghi2011, Nateghi2014} use a large data set representing the Gulf Coast region of the U.S. This is one of the largest power outage data sets in the literature. The data set includes estimates of wind gust speed, duration of wind speed exceeding 20 m/s, land cover, soil moisture, antecedent precipitation and mean annual precipitation. Power system data includes numbers of poles, transformers and switches, length of overhead and underground lines and number of impacted customers. These data are mapped to 3.66 km by 2.4 km grid cells. Restoration data are available for three hurricane events. 

Mitsova et al. \cite{Mitsova2018} studies Florida's power restoration after Hurricane Irma. Their data set includes many county-specific features, such as percent of customers without power by account type, urban/rural classification, number of accounts, breakdown of accounts between Investor-Owned, Rural Cooperative and Municipal Utility. Socioeconomic variables such as race/ethnicity, population density, \% renter occupied housing, \% population with less than high school education, and unemployment rate also play a key role in their analysis. Finally, the data set represents which counties had their centroid in the hurricane-force wind swath vs. tropical storm wind swath. 

One of the most commonly used data sets for modeling various aspects of disaster recovery is that used in Liu et al. \cite{Liu2007}. This data set is used for many publications, some not directly modeling lifeline restoration \cite{Liu2005}, and others extending existing restoration modeling work \cite{ReedDorothyA2008}. The data set includes outage data from three utility companies in the North Carolina area for six hurricanes and eight ice storms. The data are collected at the county level for land cover, number of customers affected, type of device affected, population density, outage start time compared to start time of first outage, estimated wind speed, seven-day rainfall and ice accretion. Reed \cite{ReedDorothyA2008} uses a subset of this data set and data from the 1999 French winter storms for their model. 

The work of Yu and Baroud \cite{Yu2019} is another that utilizes a data set from Shelby County, Tennessee. Their data set comprises outage data from fifteen storms for MLGW between 2007 and 2017. Shelby County and MLGW have provided data for research in the past that resulted in many extensive works, most notably a MCEER project in the '90s  \cite{Chang1996,ShinozukaM1998}. The data set presented in Chang et al. \cite{Chang1996} comprises layouts for water, electricity and natural gas systems, restoration data for the 1994 Northridge earthquake, utility usage data, census data and economic data.

There are several studies that aggregate data from many events worldwide to build their models, as well as studies focusing on a specific geographic area for restoration data. D{\'\i}az-Delgado Bragado \cite{diazdelgado2016} builds a database of restoration data for 31 earthquake events from around the world, 1923-2015, considering water, power, gas and telecommunications systems and uses it to fit gamma cumulative distribution functions. Monsalve and de la Llera \cite{Monsalve2019} also compile earthquake restoration data, encompassing 6 different earthquakes and various infrastructure systems. Kammouh et al. \cite{KammouhOmar2018Deaa} likewise bring together worldwide earthquake restoration data, including 32 earthquakes in their study. Zorn and Shamseldin \cite{Zorn2015} is another work that brings together restoration data from multiple events, 18 total, including earthquakes, hurricanes and other types of disasters, for electricity, water, gas and telecommunications systems. Finally, Duffey \cite{Duffey2019} collects power restoration data for 13 disaster events between 2012 and 2018 through ``power tracker" or ``outage map" websites. 

Nojima et al. \cite{nojima2002, nojima2005, nojima2012, Nojima2014} collect data sets from Japan earthquake events as the basis for their models. This includes seismic intensity from the Japan Meteorological Agency, spatially distributed population data and network vulnerability data for water and gas systems. Restoration data for electricity, water and gas systems are also used. The data sets are collected from the 1995 Hyogoken-Nambu earthquake and the 2011 T\={o}hoku  earthquake. 

MacKenzie and Barker \cite{Mackenzie2013} use publicly available U.S. outage data, collected by the U.S. Department of Energy through form OE-417, along with state population data. The data set includes duration, location (state), and cause of the outage between January 2002 to June 2009. Barker and Baroud \cite{BarkerKash2014Phmo} and Barabadi and Ayele \cite{Barabadi2018} use the same data set, while Mukherjee et al. \cite{Mukherjee2018} utilize a larger data set of OE-417 submissions, containing information from January 2000 to July 2016. They use state-level population data, climate data from the U.S. National Oceanic and Administrative Administration, electricity consumption patterns from the U.S. Energy Information Administration, Urban/Rural and Land/Water percentages from the U.S. Census Bureau and state-level economic characteristics from the U.S. Bureau of Economic Analysis. 

Bessani et al. \cite{Bessani2016} use outage data from a single Brazilian power distribution system during 2012-2015. The data set included duration and cause. Mojtahedi et al. \cite{Mojtahedi2017} use transportation restoration data from Australia 1992-2012. The data set comprises 4245 transportation projects from a variety of causes including flood, storm and bushfire, detailing cost and restoration duration. Barabadi and Ayele \cite{Barabadi2018} use two separate data sets of Iranian lifelines for their case studies (in addition to the US outage data). The first includes date, cause, number of customers affected, location of power distribution system, number of assigned recovery crews and system age/condition for 64 power outage events from 1998 to 2014. The second data set is restoration data for over 30 bridges during the period 2003 to 2015.

\subsubsection{Data Sources}

Direct collaboration with utility companies is again a common data source \cite{Chang1996,Park2006,Nateghi2011,Nateghi2014,Mitsova2018,Cooper1998C}. Nateghi et al. \cite{Nateghi2011,Nateghi2014} supplement their utility-provided data with data collected from a commercial weather forecasting service and the National Land Cover database. Mitsova et al. \cite{Mitsova2018} collect additional data from the American Community Survey for their model. Several modelers got their data sets from public U.S. government data sources \cite{Mukherjee2018,Mackenzie2013,BarkerKash2014Phmo,Barabadi2018}. The most common data source is data sets from previous studies, such as the worldwide restoration data sets in Zorn and Shamseldin \cite{Zorn2015} and Kammouh et al.\cite{KammouhOmar2018Deaa}. Using a novel approach, Duffey \cite{Duffey2019} makes use of ``outage tracker" websites to gather restoration data after multiple disasters. Sources outside the U.S. are used in a number of different works \cite{Bessani2016,Mojtahedi2017,Barabadi2018}. Finally Public outage reports are used in Duffey and Ha \cite{Duffey2013}.

\section{Discussion}\label{sec:discussion}

We identified several topics worth discussing through completing this literature review. These topics are model validation and testing, modeling restoration of interdependent systems, benchmarking testbeds, and unique data sources for recovery data. These topics have relevance to the future direction of the lifeline restoration modeling literature. 

\subsection{Validation and Testing Methods}

As a precursor to this section, we want to acknowledge that model validation is a contested concept with many definitions, recommendations, and best practices across disciplines \cite{WeinsteinMiltonC.2003PoGP,RoseKennethA2015Pbmp}. There is a tendency to think that every model was developed with the intent to predict, and thus every model should be validated using out-of-sample testing. However, there are many reasons for modeling outside of prediction \cite{epstein2008model}. Given that this paper's goal is to discuss the use of empirical quantitative data, reviewed papers use data for model calibration, validation, or application through a case study. Acknowledging that out-of-sample testing is not applicable or feasible for every modeling study, this subsection discusses what out-of-sample validation and testing techniques have been used in the field so far.

As one would expect, statistical models have the widest variety of out-of-sample validation and testing approaches. Cross-validation is used in a few models for parameter fitting or model comparison \cite{Mackenzie2013,Nateghi2014, Guikema2010,  Yu2019}. Some modelers split their data sets into training and test sets by withholding information from some disaster events \cite{Davidson2017, Nateghi2011,Liu2007}. Park et al. \cite{Park2006} fit a curve to restoration data from one event and compared the fitted parameters to that of another event.  

There are several out-of-sample validation methods used by simulation models as well. For the projects that partnered with LADWP \cite{Cagnan2007,Cagnan2006, Xu2007,Tabucchi2010,Brink2012,Brink2009, mceer080008}, this was to use restoration data from the 1994 Northridge earthquake. The authors perform the validation by setting model input parameters (e.g. number of repair crews) to be equivalent to the Northridge conditions and compare the simulated restoration time to the actual restoration time from the Northridge event. Other papers that compare their model output to restoration data included Isumi et al. \cite{Isumi1985} and He and Cha \cite{Xian2018}. A comparison between model output for a theoretical disaster event and restoration data from a similar disaster event in a different location \cite{Ramachandran2015} is one of the more inventive validation methods seen in the literature.

There are no optimization models in the reviewed literature that were tested out of sample. However, given the nature of an optimization model, this should not come as a surprise. The goal of optimization is usually to perform better than the status quo, thus the restoration time estimates from an optimization model would nearly always be below real-world restoration times. The contribution of an optimization model is typically a new model formulation \cite{Yan2009,Nurre2012} or solution approach \cite{González2016}. 

Overall, the reviewed literature encompassed a broad variety of validation and testing approaches. While the authors encourage the use of out-of-sample model evaluation, we understand that this is not always possible, nor does it always make sense. However, as models continue to become more generalizable and data more available, we hope to see more out-of-sample evaluation take place in this field. 

\subsection{Modeling Interdependent Systems}

\begin{figure}[!h]
\caption{Number of publications over time on modeling interdependent restoration.}
\centering
\includegraphics[width=1\textwidth]{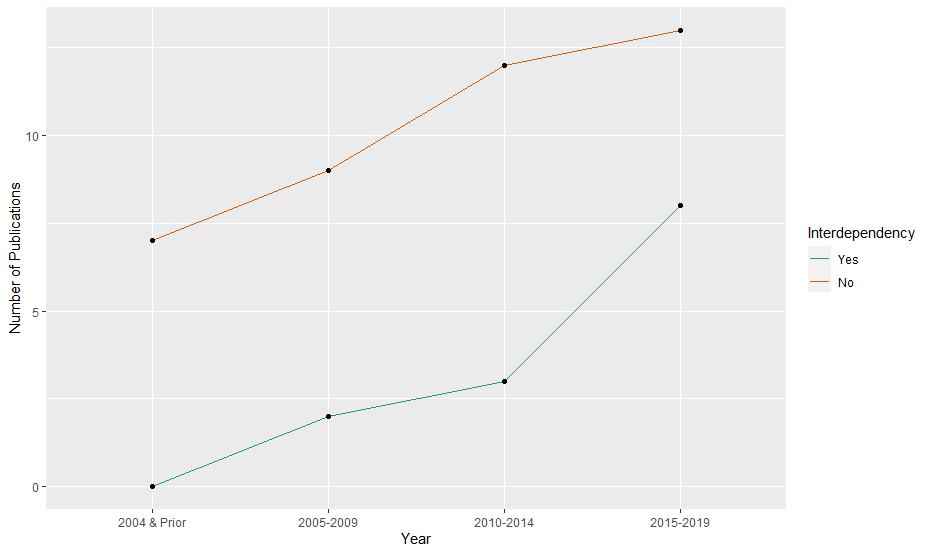}
\label{inttime}
\end{figure}

Interdependency is increasingly recognized as an important factor to consider while modeling lifeline restoration, as seen in Figure~\ref{inttime}. Lifeline systems are interdependent by nature. As an example, power generators require water for cooling and electricity needs to be available for water pumps to function. Quantifying these interdependencies regarding restoration is an ongoing challenge for modelers, but one that is actively being worked on by researchers. 

\begin{figure}[!h]
\caption{Breakdown of publications by modeling approach on interdependent restoration.}
\centering
\includegraphics[width=1\textwidth]{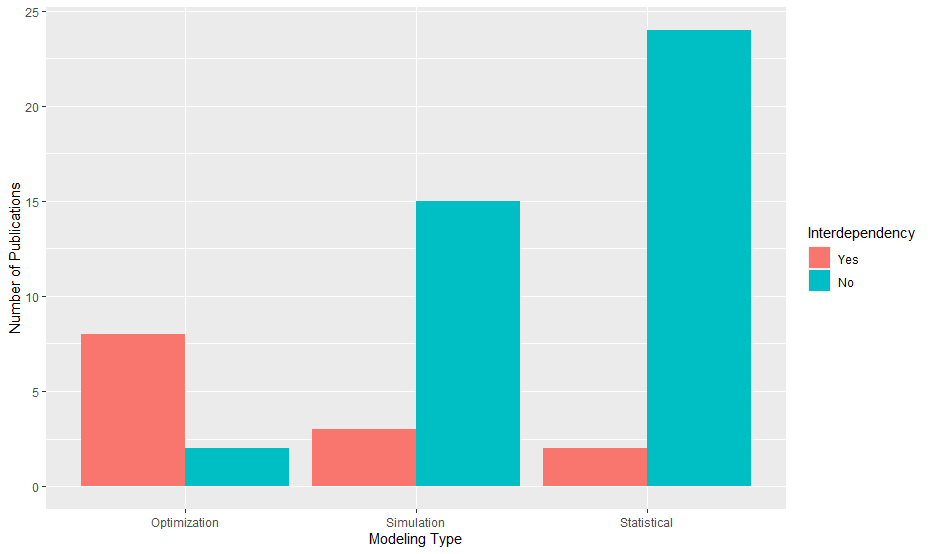}
\label{modint}
\end{figure}

There was an increase in studies of cascading failures \cite{IntCasc1,IntCasc3, IntCasc2} in recent years, and there is a broad recognition that lifelines are restored in an interdependent fashion \cite{RinaldiS.M2001Iuaa,Sharkey2016}. In contrast, our review shows that only 80\% of the reviewed literature does not consider interdependencies directly. Optimization models have the longest history of success in incorporating interdependencies in their models, as seen in Figure~\ref{modint}. The rest of this section discusses a few of the methods used to model interdependent restoration in the reviewed literature and promising approaches that, to our knowledge, have yet to be applied in a restoration modeling context. 

Lee et al. \cite{Lee2007} is the oldest instance of modeling interdependent infrastructure restoration in the reviewed literature. They consider power, telecommunications and transportation systems, and model five types of interdependency: input dependence, mutual dependence, shared dependence, exclusive-or dependence and co-located dependence. The authors include interdependencies as a constraints in their problem formulation. Cavdaroglu et al. \cite{Cavdaroglu2013} utilize the same data set, but take the added step of determining an optimal restoration sequence for the lifeline systems. Their objective is to maximize the functionality of the lifeline services over the restoration period by balancing unmet demand costs and operating costs. They also model restoration interdependencies through their model constraints.

Yan and Shih \cite{Yan2009} model transportation restoration and emergency relief distribution together. While not a model of interdependent \textit{lifeline} restoration, this work shows a way to model restoration of interdependent \textit{systems}. They use a multi-objective optimization model to minimize the length of time for restoration and subsequent relief distribution. The authors note the connection between the transportation system and the ability to distribute relief. 

MacKenzie and Barker \cite{Mackenzie2013} utilize the dynamic inoperability input-output model (DIIOM) to include interdependency in their restoration model. Interdependencies in the DIIOM are quantified using commodity flow data from the U.S. Bureau of Economic Analysis. They apply the model to estimate restoration from power outages. This is the earliest non-optimization approach to modeling restoration interdependency in the reviewed literature. He and Cha \cite{Xian2018} extend the DIIOM to calculate facility-level interdependencies as opposed to system-level interdependencies in the traditional DIIOM. This facility-level approach captures interdependencies not only \emph{across}, but also \emph{within} systems. 

Other more recent models have a variety of approaches for modeling interdependent restoration \cite{Ramachandran2015,González2017,González2016,Monsalve2019}. Monsalve and de la Llera \cite{Monsalve2019} calculate a daily restoration rate for each lifeline in their model based on the lifeline type, its interdependencies, and an additive Gaussian error term. The authors utilize a least-squares criterion that minimizes the difference between the expected value of the model and the data to estimate model parameters, including lifeline interdependencies. Their model assumes that the restoration rate of a given lifeline depends on the functionality of other lifeline systems, but not on their restoration rates.  Gonz\'{a}lez et al. \cite{González2017,González2016} define three types of interdependencies: logical, physical, cyber and geographic. They account for these interdependencies through the constraints of their optimization model. Ramachandran et al. \cite{Ramachandran2015} include interdependency in their simulation model by including constraints that some tasks cannot start until others finish, e.g. power lines cannot be repaired until the road to access those lines is free of debris. 

There is a series of papers that utilize time-series restoration data and cross-correlation functions to quantify the interdependency between two lifelines \cite{Duenas-Osorio2012,Cimellaro2014,Krishnamurthy2016}. As of yet, this method of quantifying interdependencies has not been incorporated into a restoration model, but the potential is there. We believe that it could be applied in an approach similar to that of Monsalve and de la Llera (described above) \cite{Monsalve2019}. 

\subsection{Engagement through Benchmarking Testbeds}

There has been a lot of progress over the last few years in creating benchmarking testbeds to be used for recovery models. Two examples are Customizable Artificial Community (CLARC) County, created by Loggins et al. \cite{Loggins2019CMtR}, and Centerville, created by the Center for Risk-Based Community Resilience Planning at Colorado State University \cite{centerville2018}. CLARC County is a GIS data set representing an artificial hurricane-prone community of 500,000. The data set contains demographic and geographic data typically reported for U.S. census tracts and physical locations and characteristics of components of civil and social infrastructure systems along with their interdependencies. The data set exists to support infrastructure and emergency management research without compromising potentially sensitive information. The Centerville community resilience testbed is a virtual city, representing a typical middle-class city in the Midwestern U.S. that is susceptible to tornadoes and earthquakes. Buildings, transportation systems, electric power and water systems are represented in the data set, as well as socioeconomic features based on American Community Survey data for Gavleston, Texas and income data from Fort Collins, Colorado. 

These testbeds are conducive to recovery research, as they allow for complete, albeit synthetic, data sets to be used to test and compare recovery models. The two examples provided here also show that testbeds can be constructed in a variety of different ways, ranging from being completely synthetic, to being based on empirical data from a single source or from an amalgamation of sources. The areas represented by the example testbeds are different, one being an individual city, while the other a U.S. county. No matter the construction, these testbeds can provide value as boundary objects for comparison if nothing else. Given how recent these efforts are, it is unclear if the development of testbeds affects the use and collection of empirical data. 

The difficulty of collecting extensive data sets for lifeline restoration modeling is well documented in the reviewed literature. Even Loggins et al. \cite{Loggins2019CMtR} mention an extensive data collection process they attempted for New Hanover County, NC, and how the difficulties they experienced led them to create CLARC county. The likelihood of developing complete data sets on all lifelines in a community \emph{and} lifeline restoration data from a disaster event in that community is low. Even if such a data set were to be developed, security concerns may prevent it from ever entering the public domain. This makes testbeds the logical next step for development of large-scale, highly detailed optimization and simulation models of interdependent recovery. However, this does not eliminate the need for the collection of empirical data. 

The data collection of Loggins et al. for New Hanover County informed the creation of CLARC county \cite{Loggins2019CMtR}, and Centerville \cite{centerville2018} was created from an amalgamation of several empirical sources. Data availability can and sometimes should inform modeling approaches depending on modelers' objectives, although models built with no empirical data can still provide useful insights and create new knowledge (e.g., what-if analysis, facilitation of discussion, and education). Examples of data availability informing model choice include the work done with LADWP. The authors had access to a large lifeline-specific data set, which made a detailed simulation model feasible. Another example of data availability informing modeling efforts/direction is the work of Mukherjee et al. \cite{Mukherjee2018}. The authors had access to publicly available data at the state level, making a broader statistical model possible. Having the data set publicly available means others can duplicate and extend this work. There are also many examples of ``benchmarking" in the literature where authors extend the modeling efforts of a previous work using the same data set and compare results.

\subsection{Alternative Data Sources}

There are studies that fall outside of this review's inclusion criteria that still deserve mention for their usage of data sources not seen in the reviewed literature. McDaniels and Chang characterize lifeline failure interdependencies using manual content analysis of newspapers and technical reports \cite{McdanielsTimothy2007EFfC,ChangStephanieE.2007Ifii}. In contrast, Lin et al. \cite{LinLucyH2018NLPf} make use of natural language processing to analyze newspaper stories from New Zealand after the Canterbury earthquakes with the goal of tracking long-term recovery. Doubleday et al. \cite{DoubledayAnnie2019} use daily bicycle and pedestrian activity as an indicator of disaster recovery. Chang et al. \cite{ChangStephanieE.2014TDCC} use expert elicitation to characterize lifeline resilience. Expert elicitation played an important role in statewide resilience initiatives \cite{wassc2012,osspac2013}, and in the development of the Federal Emergency Management Agency's HAZUS \cite{Hazus}. 

All of the above approaches do not rely on empirical data directly related to lifelines or lifeline restoration. In particular, approaches such as those seen in Doubleday et al. \cite{DoubledayAnnie2019} are promising because they make use of empirical quantitative data that has not been used in the restoration modeling space. If data sets of this nature can be linked to lifeline restoration data, the total amount of restoration data sets available would increase. 

Another alternative data source, Lin et al. \cite{LinLucyH2018NLPf} use natural language processing to generate recovery data from news stories about disasters. While their analysis is focused on long-term recovery, a similar approach could be used for modeling shorter-term restoration, perhaps using a different source such as Twitter data \cite{RaginiJ.Rexiline2018Bdaf,ZouLei2018MTDf,MilesScottB2014RaIf}. Expert elicitation is another method that can be used to develop restoration models. Models based on expert judgment can apply techniques such as Cooke's method \cite{CookeRogerM.1991Eiu:} to create a systematic approach for eliciting expert knowledge when empirical data are unavailable or inaccessible. 

\section{A Data Management Methodology for Reproducibility in Disaster Recovery Modeling}\label{sec:reproducibility}

The research community benefits from reproducibility, which is fostered by detailed metadata and data publication. Within the analyzed publications, data descriptions are frequently lacking and data sets are rarely published. Although it is not always possible to publish data sets for a wide variety of reasons (e.g., security and privacy), this reduces the reproducibility of any research using those data sets. As data becomes increasingly prevalent across research domains, there are more advocates for increased accessibility of data sets. Gentleman and Lang \cite{ReproducibilityCompendiumsGentlemanRobert2007} go so far as to call for the publication of ``reproducible research compendiums" which include the final paper, as well as the data set, software and any other items necessary to reproduce the research. They acknowledge that this is not feasible for all research, but maintain that publishing as much information as possible is worthwhile. 

Gonz\'{a}lez-Barahona and Robles \cite{ReproducibilitySoftwareGonzález-Barahona} discuss reproducibility of empirical software engineering studies and identify elements of said studies with an impact on reproducibility. We adapt the ideas presented by Gonz\'{a}lez-Barahona and Robles to fit the disaster recovery modeling domain. It is our recommendation that all disaster recovery modeling papers using empirical data include a data description section, with at least the following components:

\begin{enumerate}
    \item Data source(s). Where did the data set come from? This should be as specific as possible. Even if the only thing an author can share is ``a certain utility company from the U.S. Southeast", that is still worthwhile for a reader to know. Where possible, links or citations to the original data source(s) should be included. 
    \item Retrieval method. How was the data set collected from the source? Examples include downloading a CSV or GIS file from a government website, receiving data via email, and using a web scraper. 
    \item Raw data set. Can the data set be shared or is it publicly available? If yes, it is recommended to provide a link to an online repository or email address for an appropriate contact about the data set along with a description of the format and features of the data set. 
    \item Data processing. What transformations were performed on the data set to get it into a usable form? What are the form and features of the processed data? Processed data sets should be stored separately from the raw data set in an online repository with a link for others to access. Obtaining a Digital Object Identifier (DOI) for the data set is a best practice after publishing it on an online repository or in a peer-reviewed journal (e.g., \cite{MUKHERJEE20182079}). Version controlled documentation can track problems in the data sets as they are found and corrected.
    \item Processed data set summary statistics. If the data set is numerical, statistics can be presented in a table. If the data set is only a network representation of an infrastructure system, a graphical representation may be sufficient. If this information cannot be shared, it can be clearly stated with a reason (e.g., national security).
\end{enumerate}

\begin{figure}[!h]
\caption{Graphical representation of the proposed data management methodology.}
\centering
\includegraphics[width=0.7\textwidth]{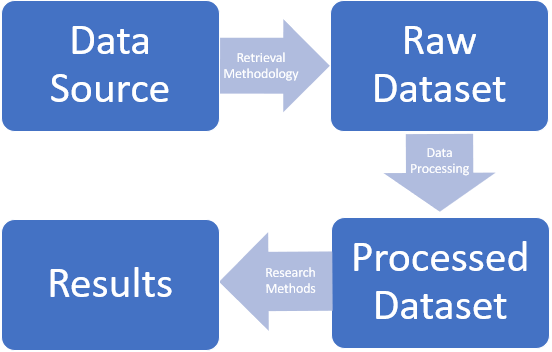}
\label{datad}
\end{figure}

Figure~\ref{datad} outlines the data management methodology.

One example of a brief, comprehensive data description in an academic publication is from Yan and Shih \cite{Yan2009}:   
``The roadway-network information includes the roadway segments and intersections in Nantou County, the location of repair points and work stations and the location of supply and demand points. The emergency repair resources include the work teams for each station and the average time required for a work team to repair each repair point. Note that, in practice, when scheduling the roadway repairs, information on the time needed to repair every repair point is given by the engineers. Several engineers estimate the repair time in advance based on experience and the level of damage. Decision-makers then use the average time (as was done in this research) to set the repair time. ... There are 46 intersections, 24 repair points, 8 demand points, 9 work stations, 24 work teams, 5 distribution centers, and 196 time unit lengths (3 days is the time length, with a time unit of 15 min), in the tests."

This description does not contain all the elements in our proposed data management methodology, but it shows what can be done with a small amount of space in a research publication. The authors do mention a specific data source in their acknowledgements section. 

We recognize that it is not always possible to share all the information in our proposed methodology due to privacy or security concerns. Even under this constraint, it is still important to make clear data management practices for research. If the data set is private, one can still provide a summary of what it contains, within the limitations of the data set provider. This creates an opportunity for future research to collect a different data set with the same features and apply the method used in the original paper. Overall, there are many opportunities for increased data sharing and higher standards for reproducibility in the field of disaster recovery modeling. The proposed method is a start to building a community of practice around data management in disaster recovery modeling. One great resource for building this community of practice is the National Science Foundation's Natural Hazards Engineering Research Infrastructure DesignSafe-CI \cite{DesignSafe}. DesignSafe-CI is a cyberinfrastructure environment for research in natural hazards engineering. The features of this cyberinfrastructure include data sharing and publication, integrated data analysis tools, high performance computing access and collaboration tools. Making use of resources like DesignSafe-CI is one way to make the disaster recovery modeling community of practice a reality. 

\section{Conclusion}

The data sources and data features used by lifeline restoration modelers vary across modeling approaches and there is no uniform methodology for how to utilize data in this research domain. This paper highlights a myriad of data sets that have been used in the past to model lifeline restoration to help build a community of practice within the broader field of disaster recovery modeling. We propose a set of best practices for managing and writing about data sets used for disaster recovery modeling.

Our review shows that direct collaboration with lifelines and publicly available data, usually from the government, were two of the most common data sources in the literature. Data sets are frequently re-used over time to provide additional insights with new/updated modeling approaches. We discuss the usage of benchmarking testbeds as an alternative way to develop and test recovery models where relevant data sets are unavailable. Expert elicitation and large text data sets are identified as additional alternative data sources. Overall, this review demonstrates the broad variety of data sources available to modelers.

Our intent for the proposed data management practices is that they will cause more data sets to become publicly available. This will encourage modelers without much experience in the disaster and hazard research to enter the research domain and open doors for disaster and hazard researchers to build models with more data than they previously had access to. With more and more data available, the goal of a generalizable model of interdependent restoration could come into view, with communities around the world as the beneficiaries.

\section*{Acknowledgements}
This work was supported by the National Science Foundation (NSF grant CMMI-1824681).

\section*{References}

\bibliography{emp_bib}

\end{document}